\documentclass[10pt,twocolumn,nofootinbib,showpacs,
aps,prd]{revtex4-1}
\usepackage{CJK}
\usepackage{amsfonts,amsmath,amssymb,mathrsfs}
\usepackage{mathtools,mathptmx,array}
\usepackage[colorlinks=true,linkcolor=blue,citecolor=blue,urlcolor=blue]{hyperref}

\def\be{\begin{equation}}
\def\ee{\end{equation}}
\def\bea{\begin{eqnarray}}
\def\eea{\end{eqnarray}}
\def\nn{\nonumber}
\def\p{\partial}
\def\cA{\mathcal{A}}

\begin{document}

\title{Are ultraspinning Kerr-Sen-AdS$_4$ black holes always superentropic?}

\author{Di Wu$^{1,2}$}
\email{wdcwnu@163.com}

\author{Puxun Wu$^{1}$}
\email{pxwu@hunnu.edu.cn}

\author{Hongwei Yu$^{1}$}
\email{hwyu@hunnu.edu.cn}

\author{Shuang-Qing Wu$^{2}$}
\email{sqwu@cwnu.edu.cn}

\affiliation{$^{1}$Department of Physics and Synergetic Innovation Center for
Quantum Effect and Applications, Hunan Normal University, Changsha, Hunan 410081,
People's Republic of China \\
$^{2}$College of Physics and Space Science, China West Normal University, Nanchong,
Sichuan 637002, People's Republic of China}

\date{\today}

\begin{abstract}
We study thermodynamics of the four-dimensional Kerr-Sen-AdS black hole and its
ultraspinning counterpart and verify that both black holes fulfil the first
law and Bekenstein-Smarr mass formulas of black hole thermodynamics. Furthermore,
we derive new Christodoulou-Ruffini-like squared-mass formulas for the usual and
ultraspinning Kerr-Sen-AdS$_4$ solutions. We show that this ultraspinning
Kerr-Sen-AdS$_4$ black hole does not always violate the reverse isoperimetric
inequality (RII) since the value of the isoperimetric ratio can be larger/smaller
than, or equal to unity, depending upon where the solution parameters lie in the
parameters space. This property is obviously different from that of the
Kerr-Newman-AdS$_4$ superentropic black hole, which always strictly violates
the RII, although both of them have some similar properties in other aspects,
such as horizon geometry and conformal boundary. In addition, it is found that
while there exists the same lower bound on mass ($m_e \geqslant 8l/\sqrt{27}$
with $l$ being the cosmological scale) both for the extremal ultraspinning
Kerr-Sen-AdS$_4$ black hole and for the extremal superentropic Kerr-Newman-AdS$_4$
case, the former has a maximal horizon radius $r_{\rm\, HP} = l/\sqrt{3}$, which is
the minimum of the latter. Therefore, these two different kinds of four-dimensional
ultraspinning charged AdS black holes exhibit some significant physical differences.
\end{abstract}


\maketitle

\section{Introduction}

Black hole is one of the most remarkable and fascinating objects in nature. It has
an event horizon beyond which any event inside has no effect. As for the horizon
topology of black holes, Hawking proved that for four-dimensional asymptotically flat,
stationary black holes satisfying the dominant energy condition, their two-dimensional
event horizon cross sections have a topology of $S^2$ sphere \cite{CMP25-152}. To
obtain different horizon topologies, one needs to relax some assumptions made in
Hawking's uniqueness theorem. Among various different possibilities, one is to
consider higher dimensional spacetimes. For example, in the five-dimensional
asymptotically flat spacetimes, apart from the well-known black hole \cite{AP172-304}
which has the horizon topology of a round $S^3$ sphere, the black ring \cite{PRL88-101101}
owns the $S^2 \times S^1$ horizon topology, while a rotating black lens solution
\cite{PRD78-064062} has the horizon topology of a lens-space $L(n,1)$. On the other
hand, if the spacetime is considered to be asymptotically nonflat, it has been found
that especially for the four-dimensional anti-de Sitter (AdS) spacetime, the Einstein
equation also admits topological solutions with their event horizons being Riemann
surfaces of any genus, namely, planar, toroidal and hyperbolic horizons \cite{PLA201-27,
PRD54-3840,PRD56-3600,PRD56-6475,PLB353-46,PRD54-4891,AIPS13-311,PRD57-6127,PRD66-064022,
PRD92-044058,PRD93-124028,NPB545-434,CQG17-2783,NPB600-272}, and rotating black string
solutions in all dimensions \cite{CQG20-2827}. Higher dimensional spacetimes can have
even more rich horizon topologies; for instance, the event horizons of the d-dimensional
asymptotically AdS black holes are also possible to be the Einstein manifolds with
positive, zero, or negative curvature \cite{CQG16-1197}.

Recently, a new class of AdS black holes \cite{PRL115-031101,JHEP0114127,PRD89-084007},
which is considered as ultraspinning since one of their rotation angular velocities is
boosted to the speed of light, has received considerable interest and enthusiasm. This
kind of black hole, occasionally called the ``black spindle" spacetime \cite{1912.03974}
because of its bottle-shaped horizon \cite{PRD93-124028}, has a noncompact horizon topology
since its spherical horizon has two punctures at the north and south poles, although
it has a finite horizon area. The ultraspinning black hole violates the reverse isoperimetric inequality (``RII") \cite{PRD84-024037,PRD87-104017}, which implies that the Schwarzschild-AdS
black hole has a maximum upper entropy. Due to the fact that the ultraspinning black hole can
exceed the maximum entropy bound, it is also dubbed ``superentropic". Remarkably,
it has been shown \cite{PRL115-031101} that the superentropic black hole solution
can be alternatively obtained by taking a simple ultraspinning limit from the usual
rotating AdS one. This solution generating procedure is very simple: first recast
the rotating AdS black hole in the frame rotating at infinity, then boost one rotation
angular velocity to the velocity of light, and finally compactify the corresponding
azimuthal direction. Up to date, a lot of new superentropic black hole solutions
\cite{JHEP0615096,JHEP0816148,PRD95-046002,1702.03448,JHEP0118042} from the known
rotating AdS black holes have been obtained so far. Very recently, it has been found
that the superentropic black hole can also be obtained by running a conical deficit
from the usual rotating AdS black hole \cite{JHEP0220195}. In addition, other aspects
of the superentropic black holes, including thermodynamic properties \cite{PRL115-031101,
JHEP0615096,PRD95-046002,1702.03448,JHEP0118042,MPLA35-2050098,PRD101-024057,PLB807-135529},
horizon geometry \cite{PRD89-084007,JHEP0615096,PRD95-046002}, geodesic motion
\cite{1912.03974}, Kerr/CFT correspondence \cite{PRD95-046002,1702.03448,JHEP0816148},
and so on, have been investigated consequently.

Although there has been much progress in the last few years in constructing superentropic
black hole solutions and studying their physical properties, ultraspinning black
holes in gauged supergravities remain to be the virgin territory and thus need to
be explored deeply, which motivates us to conduct the present work. Since the most
famous rotating charged black hole in the four-dimensional low-energy heterotic
string theory is the Kerr-Sen solution \cite{PRL69-1006}, we first consider its
generalization by including a nonzero negative cosmological constant, namely the
Kerr-Sen-AdS$_4$ black hole, and then get its ultraspinning counterpart. Along the
way, we also address their thermodynamical properties and show that the obtained
thermodynamical quantities perfectly obey both the extended first law and the
Bekenstein-Smarr mass formulas for both black holes.

The remaining part of this paper is organized as follows. In Sec. \ref{II}, we make
a recapitulation about the Kerr-Sen black hole, and then turn to the Kerr-Sen-AdS$_4$
black hole solution in the four-dimensional gauged Einstein-Maxwell-dilaton-axion
(EMDA) theory and investigate its thermodynamics. In Sec. \ref{III}, after the
ultraspinning Kerr-Sen-AdS$_4$ black hole solution is constructed, its horizon
topology and conformal boundary, thermodynamical properties, bounds on the mass and
horizon radius of extremal ultraspinning charged AdS$_4$ solutions, and the RII are
subsequently discussed. In doing so, we establish novel Christodoulou-Ruffini-like
squared-mass formulas for the Kerr-Sen-AdS$_4$ black hole and its ultraspinning
counterpart. Differentiating this formula with respect to its individual
thermodynamical variable gives the expected thermodynamical quantities which
satisfy both the first law and the Bekenstein-Smarr mass formula without applying
the chirality condition ($J = Ml$). Then, we discuss the reduced form of the mass
formulas after imposing the chirality condition. We demonstrate that the RII does
not always hold true for this ultraspinning Kerr-Sen-AdS$_4$ black hole, since
the value of the isoperimetric ratio can be larger/smaller than, or equal to unity,
depending upon the range of the solution parameters. This marks a striking difference
from the Kerr-Newman-AdS$_4$ superentropic black hole. Finally, we end up with our
summaries in Sec. \ref{IV}.

\section{Kerr-Sen black hole and its AdS version}\label{II}

\subsection{A brief review of Kerr-Sen black hole}

By using a solution generating technique with the Kerr black hole as the seed
solution, Sen \cite{PRL69-1006} obtained a new solution of the four-dimensional
rotating charged black hole, which is named as the Kerr-Sen black hole. It is
an exact solution to the four-dimensional low-energy heterotic string theory,
also known as the EMDA theory, whose Lagrangian has two different but completely
equivalent forms,
\bea
\mathcal{L} &=& \sqrt{-g}\Big[R - \frac{1}{2}(\p\phi)^2
 -e^{-\phi}F^2 -\frac{1}{12}e^{-2\phi}H^2 \Big] \\
&=& \sqrt{-g}\Big[R -\frac{1}{2}(\p\phi)^2 -\frac{1}{2}e^{2\phi}(\p\chi)^2
  -e^{-\phi}F^2 \Big] \nn \\
&& +\frac{\chi}{2}\epsilon^{\mu\nu\rho\lambda}F_{\mu\nu}F_{\rho\lambda} \, ,
\eea
where $R$ is the Ricci scalar, $\phi$ is the dilaton scalar field, $F_{\mu\nu}$
is the Faraday-Maxwell electromagnetic tensor and $F^2 = F_{\mu\nu}F^{\mu\nu}$,
$\chi$ is the axion pseudoscalar field dual to the three-form antisymmetric tensor:
$H = -e^{2\phi}\,{\star}d\chi$, and $H^2 = H_{\mu\nu\rho}H^{\mu\nu\rho}$, and
$\epsilon^{\mu\nu\rho\lambda}$ is the four-dimensional Levi-Civita antisymmetric
tensor density.

The Kerr-Sen black hole solution can be expressed in the Boyer-Lindquist coordinates
as \cite{PTP92-47,JMP44-1084}
\be\label{KS}
\begin{split}
ds^2 &= -\frac{\Delta}{\Sigma}\big(dt -a\sin^2\theta\, d\varphi\big)^2
 +\frac{\Sigma}{\Delta}dr^2 +\Sigma d\theta^2 \\
&\quad +\frac{\sin^2\theta}{\Sigma}\big[adt -(r^2 +2br +a^2)d\varphi\big]^2 \, , \\
A &= \frac{qr}{\Sigma}\big(dt -a\sin^2\theta\, d\varphi\big) \, ,\\
\phi &= \ln\Big(\frac{r^2 +a^2\cos^2\theta}{\Sigma}\Big)\, , \quad
\chi = \frac{2ba\cos\theta}{r^2 +a^2\cos^2\theta} \, ,
\end{split}
\ee
where
\bea
\Delta = r^2 +2(b -m)r +a^2 \, , \quad \Sigma = r^2 +2br +a^2\cos^2\theta \, ,  \nn
\eea
in which $b = q^2 /2m$ is the dilatonic scalar charge, the parameters $m$ and $q$ are
the mass and electric charge of the black hole, respectively, and its angular momentum
is $J = ma$.

Although the metric and gauge field of the Kerr-Sen black hole have almost the same
forms as those of the Kerr-Newman black hole (therefore they own many very similar
physical properties, such as geometric feature \cite{PTP92-47}, quantum thermal
property and thermodynamical four laws \cite{JMP44-1084}, instability of the bound
state of the charged mass scalar field and CFT$_2$ holographic duality of the
scattering process, etc.), there are some significant differences between them.
For example, the Kerr-Sen black hole is a nonvacuum, nonalgebraically special
solution in the four-dimensional low-energy heterotic string theory. In addition
to the metric and an Abelian vector field, the Kerr-Sen solution contains another
two nongravitational fields: an antisymmetric third-order tensor field (or a dual
axion pseudoscalar field), and a dilaton scalar field, which are absent from the
Kerr-Newman solution. As for the Petrov-Pirani classification, the Kerr-Newman
black hole, which is an exact electric vacuum solution to the Einstein-Maxwell
theory, belongs to the family of type-D, while the Kerr-Sen solution is of type-I
\cite{PRD52-5826}. While the electrostatic potentials of the stringy left- and
right-movers of the Kerr-Sen black hole are identical, they are unequal in the
Kerr-Newman case \cite{PLB608-251}. Of course, there are still many other salient
different aspects between them, for instance the capture region of scattered
photons, the emission probability of black hole evaporation process, the magnetic
induction ratio, and so on (see Refs. \cite{CQG30-135005,PRD93-064028} and
references therein).

\subsection{Kerr-Sen-AdS$_4$ black hole solution}

We now turn to include a nonzero negative cosmological constant into the Kerr-Sen
solution and present a simple form of the Kerr-Sen-AdS$_4$ black hole, which is an
exact solution to the gauged EMDA theory, whose Lagrangian has the following form
\bea
\bar{\mathcal{L}} &=& \sqrt{-g}\Big\{R - \frac{1}{2}(\p\bar{\phi})^2
 -\frac{1}{2}e^{2\bar{\phi}}(\p\bar{\chi})^2 -e^{-\bar{\phi}}F^2 \nn \\
&& +\frac{1}{l^2}\big[4 +e^{-\bar{\phi}} +e^{\bar{\phi}}(1 +\bar{\chi}^2)\big] \Big\}
 +\frac{\bar{\chi}}{2}\epsilon^{\mu\nu\rho\lambda}F_{\mu\nu}F_{\rho\lambda} \, ,
\quad
\eea
with $l$ being the cosmological scale. Distinct from the ungauged case, now the above
Lagrangian receives a potential term contributed from the dilation and axion fields,
so it is impossible to reexpress it in the dualized version in terms of the three-form
field that appeared in the ungauged Lagrangian.

Written in terms of the Boyer-Lindquist coordinates and adapted to the frame rotating
at infinity, the Kerr-Sen-AdS$_4$ black hole solution can be given by the following
exquisite forms:
\be\label{KSAdS}
\begin{split}
d\bar{s}^2 &= -\frac{\Delta_r}{\Sigma}\Big(dt -\frac{a\sin^2\theta}{\Xi}d\varphi\Big)^2
 +\frac{\Sigma}{\Delta_r}dr^2 +\frac{\Sigma}{\Delta_\theta}d\theta^2  \\
&\quad +\frac{\Delta_\theta\sin^2\theta}{\Sigma}\Big(adt
 -\frac{r^2+2br+a^2}{\Xi}d\varphi\Big)^2 \, , \\
\bar{A} &= \frac{qr}{\Sigma}\Big(dt -\frac{a\sin^2\theta}{\Xi}d\varphi\Big) \, ,  \\
\bar{\phi} &= \ln\Big(\frac{r^2 +a^2\cos^2\theta}{\Sigma}\Big) \, , \quad
\bar{\chi} = \frac{2ba\cos\theta}{r^2 +a^2\cos^2\theta} \, ,
\end{split}
\ee
where $\Sigma = r^2 +2br +a^2\cos^2\theta$ as before, and now we have
\bea
\Delta_r &=& \Big(1 +\frac{r^2 +2br}{l^2}\Big)(r^2 +2br +a^2) -2mr \, , \nn \\
\Delta_\theta &=& 1 -\frac{a^2}{l^2}\cos^2\theta \, , \quad
\Xi = 1 -\frac{a^2}{l^2} \, . \nn
\eea
Obviously, the above solution (\ref{KSAdS}) consistently reduces to the Kerr-Sen
black hole solution (\ref{KS}) when the AdS radius $l$ tends to infinity.

It should be pointed out that more general solutions have been already constructed
\cite{NPB717-246,PRD89-065003} in the special case of the pairwise equal charge
parameters of the four-dimensional gauged STU supergravity theory. As the gauged
EMDA theory is a more special case of that theory, therefore the above solution can
be included as a special case obtained in \cite{NPB717-246,PRD89-065003}; however,
here we present it in a slightly different and more suitable form.

\subsection{Thermodynamics}

Now we are in a position to investigate thermodynamics of the Kerr-Sen-AdS$_4$ black
hole. In the framework of the extended phase space \cite{CPL23-1096,CQG26-195011} (see
Ref. \cite{CQG34-063001} for a fairly comprehensive review), thermodynamic quantities
associated with the above solution (\ref{KSAdS}) can be computed through the standard
method and have the following expressions:
\be\label{Therm}
\begin{split}
&\bar{M} = \frac{m}{\Xi} \, , \quad \bar{J} = \frac{ma}{\Xi^2} \, , \quad
\bar{Q} = \frac{q}{\Xi} \, , \\
&\bar{T} = \frac{(r_+ +b)(2r_+^2 +4br_+ +l^2 +a^2)
 -ml^2}{2\pi(r_+^2 +2br_+ +a^2)l^2} \, , \\
&\bar{S} = \frac{\pi(r_+^2 +2br_+ +a^2)}{\Xi} \, , \quad
 \bar{\Omega} = \frac{a\Xi}{r_+^2 +2br_+ +a^2} \, , \\
&\bar{\Phi} = \frac{qr_+}{r_+^2 +2br_+ +a^2} \, .
\end{split}
\ee

It is easy to very that these thermodynamic quantities (\ref{Therm}) satisfy the
Bekenstein-Smarr mass formulas
\be
\bar{M} = 2\bar{T}\bar{S} +2\bar{\Omega}\bar{J} +\bar{\Phi}\bar{Q} -2\bar{V}\bar{P} \, .
\ee
Here $\bar{V}$ is the thermodynamic volume
\be
\bar{V} = \frac{4\pi}{3\Xi}(r_+ +b)(r_+^2 +2br_+ +a^2) \, ,
\ee
which is conjugate to the pressure $\bar{P} = 3/(8\pi\,l^2)$. Unfortunately, the first
law, however, boils down to a differential identity only
\be
d\bar{M} = \bar{T}d\bar{S} +\bar{\Omega}d\bar{J} +\bar{\Phi}d\bar{Q}
 +\bar{V}d\bar{P} +\bar{J}d\Xi/(2a) \, .
\ee
The reason for this is simply because we have just adopted the rotating frame at infinity,
not the rest frame at infinity.

The transformation of the above Kerr-Sen-AdS$_4$ solution into the frame rest at infinity
can be easily done by taking a simple coordinate transformation $\varphi\rightarrow
\varphi -a\,t/l^2$. After a tedious calculation of the thermodynamic quantities in
this rest frame, it is not difficult to find that only the mass, the angular velocity
and the thermodynamic volume are different from those given in Eq. (\ref{Therm})
and can be written as follows:
\be
\widetilde{M} = \bar{M} +\frac{a}{l^2}\bar{J} \, , \quad
 \widetilde{\Omega} = \bar{\Omega} +\frac{a}{l^2} \, ,
\quad \widetilde{V} = \bar{V} +\frac{4\pi}{3}a\bar{J} \, .
\ee
In this situation, now thermodynamic quantities can indeed satisfy both the standard
forms of the first law and the Bekenstein-Smarr mass formula simultaneously,
\be
\begin{split}
d\widetilde{M} &= \bar{T}d\bar{S} +\widetilde{\Omega} d\bar{J}
 +\bar{\Phi}d\bar{Q} +\widetilde{V}d\bar{P} \, , \\
\widetilde{M} &= 2\bar{T}\bar{S} +2\widetilde{\Omega}\bar{J}
 +\bar{\Phi}\bar{Q} -2\widetilde{V}\bar{P} \, .
\end{split}
\ee
It is not difficult to check that the above differential and integral mass formulas
can be deduced from the following Christodoulou-Ruffini-like squared-mass formulas:
\be
\widetilde{M}^2 = \Big(1 +\frac{8\bar{P}\bar{S}}{3}\Big)\bigg[
 \Big(1 +\frac{8\bar{P}\bar{S}}{3}\Big)\frac{\bar{S}}{4\pi}
 +\frac{\pi\,\bar{J}^2}{S} +\frac{\bar{Q}^2}{2}\bigg] \, .
\ee

\section{Ultraspinning Kerr-Sen-AdS$_4$ black hole}\label{III}

\subsection{The ultraspinning solution}

To construct the ultraspinning version from the above Kerr-Sen-AdS$_4$ black hole
solution (\ref{KSAdS}), we just need to perform three steps: (i) redefine the angle
coordinate $\varphi$ by multiplying it with a factor $\Xi$; (ii) take the $a \to l$
limit; and (iii) then compactify the $\varphi$ direction with a period of the
dimensionless parameter $\mu$. Having finished these steps, we then obtain the
ultraspinning Kerr-Sen-AdS$_4$ black hole solution,
\be\label{SEKSAdS}
\begin{split}
d\hat{s}^2 &= -\frac{\hat{\Delta}_r}{\hat{\Sigma}}\big(dt
 -l\sin^2\theta\, d\varphi\big)^2 +\frac{\hat{\Sigma}}{\hat{\Delta}_r} dr^2
 +\frac{\hat{\Sigma}}{\sin^2\theta} d\theta^2 \\
&\quad +\frac{\sin^4\theta}{\hat{\Sigma}}\big[ldt
 -(r^2 +2br +l^2)d\varphi\big]^2 \, , \\
\hat{A} &= \frac{qr}{\hat{\Sigma}}\big(dt -l\sin^2\theta\, d\varphi\big) \, , \\
\hat{\phi} &= \ln\Big(\frac{r^2 +l^2\cos^2\theta}{\hat{\Sigma}}\Big) \, , \quad
\hat{\chi} = \frac{2bl\cos\theta}{r^2 +l^2\cos^2\theta} \, ,
\end{split}
\ee
where
\bea
\hat{\Delta}_r = (r^2 +2br +l^2)^2l^{-2} -2mr \, , \quad
\hat{\Sigma} = r^2  +2br +l^2\cos^2\theta\, . \nn
\eea

\subsection{Horizon geometry and conformal boundary}

In this subsection, we first focus on other basic properties, such as the horizon
geometry and conformal boundary of the ultraspinning Kerr-Sen-AdS$_4$ black hole.
To ensure that the geometry is free of any closed timelike curve (CTC), one needs
to check whether the inequality $g_{\varphi\varphi} \geqslant 0$ satisfies or not.
From its expression $g_{\varphi\varphi} = \frac{2mrl^2}{\hat{\Sigma}}\sin^4\theta$,
we find that $g_{\varphi\varphi}$ is always positive (due to $m\geqslant 0$,
$r\geqslant 0$ and $\hat{\Sigma}\geqslant 0$) in the entire spacetime and thus the
spacetime is free of CTC.

To investigate the geometry of the event horizon, we consider the constant ($t, r$)
surface on which the induced metric reads
\be\label{hm}
d\hat{s}^2_h = \frac{\hat{\Sigma}_+}{\sin^2\theta}d\theta^2
 +\frac{2mr_+l^2}{\hat{\Sigma}_+}\sin^4\theta d\varphi^2 \, ,
\ee
where $\hat{\Sigma}_+ = \hat{\Sigma}|_{r_+}$. This metric appears to be singular at
$\theta = 0$ and $\theta = \pi$. To examine whether the metric is ill-defined at
$\theta = 0$ and $\theta = \pi$, one can analyze it in two limits: $\theta\rightarrow 0$
and $\theta\rightarrow \pi$. As an example, let us consider the small angle case
$\theta \sim 0$, (similarly for the $\theta \sim \pi$ case). By introducing a new
variable $k = l(1 -\cos\theta)$ for a small angle $\theta$, and noting that
$\sin^2\theta \simeq 2k/l$, the two-dimensional cross section (\ref{hm}) for small
$k$ becomes
\be
d\hat{s}^2_h = \big(r_+^2 +2br_+ +l^2\big)\Big(\frac{dk^2}{4k^2}
 +\frac{4k^2}{l^2}d\varphi^2 \Big) \, ,
\ee
which is clearly a metric of constant, negative curvature on a quotient of the hyperbolic
space $\mathbb{H}^2$. This result is very similar to that of the Kerr-Newman-AdS$_4$
superentropic black hole \cite{JHEP0615096}. Due to the symmetry, the $\theta = \pi$
limit gives rises to the same result. Apparently, the space is free from pathologies
near the north and south poles. Topologically, the event horizon is a sphere with two
punctures, and occasionally is called as the black spindle. This implies that the
above ultraspinning Kerr-Sen-AdS$_4$ black hole enjoys a finite area but noncompact
horizon.

Next, we want to investigate the conformal boundary of the ultraspinning Kerr-Sen-AdS$_4$
black hole. Multiplying the metric (\ref{SEKSAdS}) with the conformal factor $l^2/r^2$
and taking the $r\to\infty$ limit, we find that the boundary metric has the form
\be\label{bdry}
ds_{bdry}^2 = -dt^2 +2l\sin^2\theta\, dtd\varphi +l^2d\theta^2/{\sin^2\theta} \, ,
\ee
which is the same one as that of the superentropic Kerr-Newman-AdS$_4$ black hole
\cite{JHEP0615096}. It is easy to see that the coordinate $\varphi$ is null on the
conformal boundary. In the small $k = l(1 -\cos\theta)$ limit, we can reexpress the
conformal boundary metric (\ref{bdry}) as
\be
ds_{bdry}^2 = -dt^2 +4kdtd\varphi +dk^2/(4k^2) \, ,
\ee
which can be interpreted as an AdS$_3$ written as a Hopf-like fibration over $\mathbb{H}^2$.
It means again that the metric has nothing pathological near two poles $\theta = 0$ and
$\theta = \pi$.

\subsection{Mass formulas}

Now we shall investigate thermodynamics of the ultraspinning Kerr-Sen-AdS$_4$ black hole.
Its fundamental thermodynamic quantities can be obtained through the standard method and
are given below
\bea\label{therm}
&& M = \frac{\mu m}{2\pi} \, , \quad J = \frac{\mu ml}{2\pi} \, , \quad
 Q = \frac{\mu q}{2\pi} \, , \nn \\
&& T = \frac{r_+ +b}{\pi l^2} -\frac{m}{2\pi(r_+^2 +2br_+ +l^2)}
 = \frac{3r_+^2 +2br_+ -l^2}{4\pi\, r_+l^2} \, , \nn \\
&& S = \frac{1}{2}\mu\big(r_+^2 +2br_+ +l^2\big) \, , \quad
 \Omega = \frac{l}{r_+^2 +2br_+ +l^2} \, , \nn \\
&&\Phi = \frac{qr_+}{r_+^2 +2br_+ +l^2} \, .
\eea
Note that the angular momentum and the mass satisfy a chirality condition: $J = Ml$.
Usually, the mass and angular momentum are computed by the conformal completion method
\cite{PRD73-104036}. However, the angular momentum can also be evaluated correctly by
the Komar integral, while the mass can be obtained via the Komar integral after the
subtraction of a divergence arising from the zero-mass background. Also, it is worthy
to point out that the angular velocity $\Omega$ is that of the event horizon because
the ultraspinning black hole is rotating at the velocity of light at infinity.

Within the framework of the extended phase space, it can be verified that the above
thermodynamical quantities completely satisfy both the first law and the Bekenstein-Smarr
mass formula:
\bea
dM &=& TdS +\Omega\, dJ +\Phi\, dQ +VdP +K d\mu \, , \label{FL} \\
M &=& 2TS +2\Omega\, J + \Phi\, Q -2VP \label{Smarr} \, ,
\eea
with the thermodynamic volume and a new chemical potential as follows
\bea
V &=& \frac{4}{3}(r_+ +b)S = \frac{2}{3}\mu(r_+ +b)(r_+^2 +2br_+ +l^2) \, , \label{V} \\
K &=& -\frac{m(r_+^2 +2br_+ -l^2)}{4\pi(r_+^2 +2br_+ +l^2)}
= \frac{l^4 -r_+^2(r_+ +2b)^2}{8\pi\,r_+l^2} \, , \label{K}
\eea
which are conjugate to the pressure $P = 3/(8\pi\, l^2)$ and the dimensionless parameter
$\mu$, respectively.

As was done in a previous work \cite{PRD101-024057}, here we propose to establish the
following simple relations
\be\label{rel}
\begin{split}
&M = \frac{\mu\Xi\bar{M}}{2\pi} \, , \quad J = \frac{\mu\Xi^2\bar{J}}{2\pi} \, ,
\quad Q = \frac{\mu\Xi \bar{Q}}{2\pi} \, , \\
&\Omega = \frac{\bar{\Omega}}{\Xi} \, , \quad S = \frac{\mu\Xi \bar{S}}{2\pi} \, ,
\quad V = \frac{\mu\Xi \bar{V}}{2\pi} \, , \\
&T = \bar{T} \, , \quad \Phi = \bar{\Phi} \, , \quad P = \bar{P} \, ,
\end{split}
\ee
and take the ultraspinning limit $a \to l$. Then we can see that the above thermodynamic
quantities given in Eq. (\ref{therm}) for the ultraspinning Kerr-Sen-AdS$_4$ black hole
can also be obtained directly from those of its corresponding usual black hole. This
further confirms that our previous method advised in Ref. \cite{PRD101-024057} is a very
effective and convenient routine to simply obtain the expected thermodynamic quantities
of all ultraspinning black holes from those of their usual counterparts by taking the
ultraspinning limit properly.

In Ref. \cite{PRD101-024057}, we have derived a new Christodoulou-Ruffini-like squared-mass
formula for the Kerr-Newman-AdS$_4$ superentropic black hole. Here we hope to seek a
similar one for the ultraspinning Kerr-Sen-AdS$_4$ black hole. Since the event horizon
equation ($\hat{\Delta}_{r_+} = 0$) can be rewritten as
\be
S^2/(\pi\, l^2) = \mu\,Mr_+ \, ,
\ee
then after using $3/l^2 = 8\pi\,P$, we get $r_+ = 8PS^2/(3\mu\,M)$. Now, we can substitute
it into the entropy $S = \mu(r_+^2 +2br_+ +l^2)/2$ and use the chirality condition
($J = Ml$) as well as $b = \pi\,Q^2/(\mu M)$ to arrive at an identity
\be
M^2 = \frac{8PS}{3\mu}\Big(\frac{4P}{3}S^2 +\pi\, Q^2\Big) +\frac{\mu J^2}{2S} \, ,
\label{sqm}
\ee
which is the expected Christodoulou-Ruffini-like squared-mass formula for the ultraspinning
Kerr-Sen-AdS$_4$ black hole. It is interesting to note that using this squared-mass formula,
it is very convenient to study black hole chemistry and possible thermodynamical phase
transition of this ultraspinning Kerr-Sen-AdS$_4$ black hole.

Leaving aside the chirality condition ($J = Ml$), it is obvious that the thermodynamical
quantities $S, J, Q, P$ and $\mu$ in Eq. (\ref{sqm}) can be regarded formally as independent
thermodynamical variables and constitute a whole set of extensive variables for the
fundamental functional relation $M = M(S,J,Q,P,\mu)$. \footnote{It should be emphasized
that not all of them are truly independent by virtue of the existence of the chirality
condition $J = Ml$ after the rotation parameter $a$ has been set to the AdS radius $l$.
However, let us ignore this relation temporarily and take a viewpoint that they look like
completely independent of each other at this moment so that there are enough parameters
to hold fixed when performing the following partial derivative manipulations. A detailed
discussion about the impact of the chirality condition on the mass formulas is presented
in the next subsection.} As did in Refs. \cite{PRD101-024057,PLB608-251,CQG17-399,
PRD21-884}, differentiating the above squared-mass formula (\ref{sqm}) with respect to
$S,J,Q,P$, and $\mu$, respectively, yields their corresponding conjugate quantities
as expected. In doing so, one can arrive at the differential first law (\ref{FL}) and
the integral Bekenstein-Smarr relation (\ref{Smarr}), with the conjugate thermodynamic
potentials correctly given by the common Maxwell relations as follows.

Differentiation of the squared-mass formula (\ref{sqm}) with respect to the entropy $S$
leads to the conjugate Hawking temperature:
\bea
T &=& \frac{\p M}{\p S}\bigg|_{(J,Q,P,\mu)} = -\, \frac{M}{2S}
 +\frac{8P}{3\mu M}\Big(\frac{8P}{3}S^2 + \pi\, Q^2\Big) \nn \\
&=& \frac{3r_+^2 +2br_+ -l^2}{4\pi\, r_+l^2} \, ,
\eea
and the corrected angular velocity and the electrostatic potential, which are conjugate
to $J$ and $Q$, respectively, are given by
\bea
\Omega &=& \frac{\p M}{\p J}\bigg|_{(S,Q,P,\mu)} = \frac{\mu J}{2SM}
 = \frac{l}{r_+^2 +2br_+ +l^2} \, , \\
\Phi &=& \frac{\p M}{\p Q}\bigg|_{(S,J,P,\mu)} = \frac{8\pi\, PQ}{3\mu M}S
 = \frac{qr_+}{r_+^2 +2br_+ +l^2} \, . \quad
\eea
Similarly for the pressure $P$ and the dimensionless quantity $\mu$, one can get the
thermodynamical volume and a new chemical potential
\bea
V &=& \frac{\p M}{\p P}\bigg|_{(S,J,Q,\mu)}
= \frac{4S}{3\mu M}\Big(\frac{8P}{3}S^2 +\pi\, Q^2\Big) \nn \\
&=& \frac{2}{3}\mu(r_+ +b)(r_+^2 +2br_+ +l^2) \, , \\
K &=& \frac{\p M}{\p \mu}\bigg|_{(S,J,Q,P)}
= \frac{M}{2\mu} -\frac{8PS}{3\mu^2 M}\Big(\frac{4P}{3}S^2 +\pi\, Q^2 \Big) \nn \\
&=& \frac{l^4 -r_+^2(r_+ +2b)^2}{8\pi\,r_+l^2} \, .
\eea
All the above conjugate quantities reproduce those expressions previously given in Eqs.
(\ref{therm}), (\ref{V}), and (\ref{K}). Anyway, with all these conjugate variables derived
from the squared-mass formula (\ref{sqm}), the differential first law (\ref{FL}) is
trivially satisfied while the integral mass formula (\ref{Smarr}) is easily checked
to be completely obeyed too.

\subsection{Chirality condition and reduced mass formulas}

Now, let us make a careful discuss about the impact of the chirality condition ($J = Ml$)
on the thermodynamical relations of the ultraspinning Kerr-Sen-AdS$_4$ black hole. Due
to the existence of the chirality condition, three thermodynamical quantities ($M, J, P$)
are not completely independent, there exists a constraint relation among them
\be
J^2 = 3M^2/(8\pi\,P) \, , \label{const}
\ee
which means that the ultraspinning Kerr-Sen-AdS$_4$ black hole is actually a degenerate
thermodynamical system. After taking into account the chirality condition physically, the
first law (\ref{FL}) and the Bekenstein-Smarr relation (\ref{Smarr}) should be constrained
by the condition (\ref{const}), and actually depict a degenerate thermodynamical system.

Considering $J$ as a redundant variable (although it is a real measurable quantity) and
eliminating $J$ from the differential and integral mass formulas in favor of $l^2 =
3/(8\pi\,P)$, the first law (\ref{FL}) and the Bekenstein-Smarr relation (\ref{Smarr})
now reduce to the following nonstandard forms (so named for their thermodynamic quantities
cannot constitute the ordinary canonical conjugate pairs due to the existence of a factor
$(1 -\Omega\, l)$ in front of $dM$ and $M$):
\be
\begin{split}
(1 -\Omega\, l)dM &= TdS +V^{\prime} dP +\Phi\, dQ +Kd\mu \, , \\
(1 -\Omega\, l)M &= 2(TS -V^{\prime}P) +\Phi\, Q \, ,
\end{split}
\ee
where
\bea
V^{\prime} = V -\frac{J\Omega}{2P} = V -\frac{4\pi}{3}\Omega\, M\, l^3 \, . \nn
\eea
In the same way, the squared-mass formula (\ref{sqm}) degenerates to
\be\label{rMs}
M^2\Big(1 -\frac{\mu}{16\pi PS} \Big) = \frac{8PS}{3\mu}\Big(\frac{4P}{3}S^2
 +\pi\, Q^2 \Big) \, .
\ee

In doing so, one actually views the enthalpy $M$ as the fundamental functional relation
$M = M(S,Q,P,\mu)$. Similar to the strategy adopted before, the above nonstandard
differential and integral mass formulas can be derived from Eq. (\ref{rMs}) by exploiting
the standard Maxwell rule. Alternately, one perhaps prefers to eliminating $P$ instead
of $J$ via Eq. (\ref{const}). As the resulted expressions are rather complicated, we
will not present them here.

\subsection{Bounds on the mass and horizon radius of
extremal ultraspinning black holes}\label{bounds}

In the following, we would like to establish some new inequalities on the mass and
horizon radius of the extremal ultraspinning black holes. We begin with the extremal
superentropic Kerr-Newman-AdS$_4$ black hole for which $\check{\Delta}_{r} = (r^2
+l^2)^2/l^2 -2mr +q^2$. Without loss of generality, here and hereafter, we shall assume
that both the mass parameter and the AdS scale are positive.

The location of the event horizon of the extremal superentropic Kerr-Newman-AdS$_4$
black hole is determined by $\check{\Delta}_{r_e} = \check{\Delta}^{\prime}_{r_e}
= 0$, which gives
\be
m_e = \frac{2r_e(r_e^2 +l^2)}{l^2} \, , \quad
q_e^2 = \frac{(r_e^2 +l^2)(3r_e^2-l^2)}{l^2} \, .
\ee
By virtue of positiveness of $q_e^2$, it is evident that the following inequalities
hold
\be
r_e \geqslant \frac{l}{\sqrt{3}} \, , \quad m_e \geqslant \frac{8l}{3\sqrt{3}} \, ,
\ee
which means that the scale of Hawking-Page phase transition: $r_{\rm HP} = l/\sqrt{3}$
is the minimum radius of the extremal superentropic Kerr-Newman-AdS$_4$ black hole,
whose mass is bounded from the lower limit: $8l/\sqrt{27}$.

Now we turn to consider the extremal case of an ultraspinning Kerr-Sen-AdS$_4$ black hole.
Its horizon is determined by $\hat{\Delta}_{r_e} = \hat{\Delta}^{\prime}_{r_e} = 0$,
which yields
\be\label{mbound}
m_e = \frac{2(r_e +b_e)(r_e^2 +2b_er_e +l^2)}{l^2} \, , \quad
b_e = \frac{l^2 -3r_e^2}{2r_e} \, .
\ee

By the requirement: $b_e = q_e^2/(2m_e) \geqslant 0$ and also $r_e \geqslant 0$, it is
clear that we must have a distinct inequality:
\be\label{rbound}
0 \leqslant r_e \leqslant \frac{l}{\sqrt{3}} \, ,
\ee
which means that the scale of Hawking-Page phase transition $r_{\rm HP} = l/\sqrt{3}$
now becomes the maximum radius of the extremal ultraspinning Kerr-Sen-AdS$_4$ black
hole. Substituting the inequality (\ref{rbound}) into Eq. (\ref{mbound}), one can find
that the extremal mass still has the same lower bound:
\be
m_e = 2\frac{(l^2 -r_e^2)^2}{l^2r_e} \geqslant \frac{8l}{3\sqrt{3}} \, .
\ee
Therefore, although the extremal Kerr-Newman-AdS$_4$ superentropic black hole and
the extremal ultraspinning Kerr-Sen-AdS$_4$ black hole share the same lower mass
bound, the Hawking-Page phase transition scale $r_{\rm HP} = l/\sqrt{3}$ marks the
dividing crest of their horizon radii. This is a remarkable signature  to distinguish
these two ultraspinning charged AdS$_4$ black holes.

\subsection{RII}

Almost a decade ago, it is conjectured \cite{PRD84-024037} that the AdS black hole
satisfies the following RII:
\be
\mathcal{R} = \Big[\frac{(D-1)V}{\cA_{D-2}}\Big]^{1/(D-1)}
 \Big(\frac{\cA_{D-2}}{A}\Big)^{1/(D-2)} \geqslant 1 \, ,
\ee
where $\cA_{D-2} = 2\pi^{[(D-1)/2]}/\Gamma[(D-1)/2]$ is the area of the unit $(D-2)$ sphere
and $A = 4S$ is the horizon area. Equality is attained for the Schwarzschild-AdS black hole,
which implies that the Schwarzschild-AdS black hole has the maximum entropy. In other words,
it indicates that for a given entropy, the Schwarzschild-AdS black hole occupies the least
volume, and hence is most efficient in storing information.

It is straightforward to check whether the ultraspinning Kerr-Sen-AdS$_4$ black hole
satisfies this RII or not. It is readily known that the area of the unit two-dimensional
sphere is $\mathcal{A}_2 = 2\mu$, the thermodynamic volume is $V = 4(r_+ +b)S/3$, and
the horizon area is $A = 4S = 2\mu(r_+^2 +2br_+ +l^2)$. Consequently, the isoperimetric
ratio now reads
\be
\mathcal{R} = \Big(\frac{r_+ +b}{2\mu}A\Big)^{\frac{1}{3}}
 \Big(\frac{2\mu}{A}\Big)^{\frac{1}{2}}
= \Big(\frac{r_+^2 +2br_+ +b^2}{r_+^2 +2br_+ +l^2} \Big)^{\frac{1}{6}} \, . \quad
\ee
Obviously, the value range of $\mathcal{R}$ is uncertain. If $b^2 < l^2$ (namely,
$q^2 < 2ml$), then $\mathcal{R} < 1$. In this case, the ultraspinning Kerr-Sen-AdS$_4$
black hole violates the RII and is superentropic. Otherwise, if $b^2 \geqslant l^2$, one
then has $\mathcal{R} \geqslant 1$. In this situation, the ultraspinning Kerr-Sen-AdS$_4$
black hole obeys the RII and is subentropic. Since the ratio of $\mathcal{R}$ depends
upon the values of the solution parameters ($q$, $m$ and $l$), thus one can see that the
ultraspinning Kerr-Sen-AdS$_4$ black hole is not always superentropic. Only when the
parameters satisfy  $q^2 < 2ml$  does it violate the RII, while the Kerr-Newman-AdS$_4$
superentropic black hole always violates the RII \cite{PRL115-031101}. As far as this
point is concerned, the ultraspinning Kerr-Sen-AdS$_4$ black hole and Kerr-Newman-AdS$_4$
superentropic black hole exhibit yet another markedly different property. This is one of
the main results that we have obtained in this paper.

\section{Conclusions}\label{IV}

In this paper, we have studied some interesting properties of the Kerr-Sen-AdS$_4$ black
hole and in particular, its ultraspinning cousin in the four-dimensional gauged EMDA
theory. After a brief review of the famous Kerr-Sen black hole solution, we presented
its exquisite generalization to include a nonzero negative cosmological constant, namely
the Kerr-Sen-AdS$_4$ black hole. Then its ultraspinning cousin was constructed via
employing a simple $a\to l$ limit procedure. The expressions of these solutions, namely
their metric, the Abelian gauge potential, the dilaton scalar, and the axion pseudoscalar
fields are very convenient for investigating their thermodynamical properties. With
these solutions at hand, all thermodynamic quantities that can be computed through
the standard method are verified to fulfil both the differential and integral mass
formulas. Moreover, new Christodoulou-Ruffini-like squared-mass formulas are displayed
for these four-dimensional black holes, from which all expected conjugate partners are
derived via differentiating them with respect to their corresponding thermodynamic
variables and are shown to consist of the ordinary canonical conjugate pairs that
appear in the standard forms of black hole thermodynamics.

Furthermore, we adopted the method proposed in Ref. \cite{PRD101-024057} to demonstrate
that all thermodynamical quantities of the ultraspinning Kerr-Sen-AdS$_4$ black hole can
be attained via applying the same ultraspinning limit to those of their corresponding
predecessors. After that, we have made a detailed discussion about the impact of the
chirality condition on the actual thermodynamics of this ultraspinning black hole.
To some extent, these aspects are very similar to those of the Kerr-Newman-AdS$_4$
superentropic black hole.

What attracts us the most in this work is to peer whether there are some other properties
peculiar to the ultraspinning Kerr-Sen-AdS$_4$ black hole. After investigating its
horizon geometry and conformal boundary, we arrived at the conclusion that both of
them are still similar to those of the Kerr-Newman-AdS$_4$ superentropic black hole.

However, when turning to investigate the extremal ultraspinning black holes and the
RII, we indeed discovered that the ultraspinning Kerr-Sen-AdS$_4$ and superentropic
Kerr-Newman-AdS$_4$ black holes exhibit some significant physical differences.

A summary of three novel consequences obtained in this paper are listed in the following
order:
\begin{enumerate}
\item
New Christodoulou-Ruffini-like squared-mass formulas are presented both for the usual
Kerr-Sen-AdS$_4$ black hole and for its ultraspinning counterpart. To the best of our
knowledge, they do not appear in the literature before, and are useful to study black
hole chemistry and possible thermodynamical phase transition of these AdS$_4$ black holes.

\item
Remarkably, it have been found that the ultraspinning Kerr-Sen-AdS$_4$ black hole is
not always superentropic, since the RII is violated only in the space of the solution
parameters that satisfy the condition $q^2 < 2ml$. Once $q^2 \geqslant 2ml$, the
ultraspinning Kerr-Sen-AdS$_4$ black hole becomes sub-entropic. This property is in
sharp contrast with the Kerr-Newman-AdS$_4$ superentropic black hole which always
violates the RII \cite{PRL115-031101}.

\item
We have established some new inequalities on the mass and horizon radius of the extremal
ultraspinning Kerr-Sen-AdS$_4$ and extremal superentropic Kerr-Newman-AdS$_4$ black
holes. It is observed that while both extremal black holes share the same mass minimum
bound $m_e \geqslant 8l/\sqrt{27}$, the scale of Hawking-Page phase transition
$r_{\rm HP} = l/\sqrt{3}$ signals the remarkably different sizes of their extremal
event horizons. That is, the horizon radius of the extremal ultraspinning Kerr-Sen-AdS$_4$
black hole never exceeds the Hawking-Page scale $r_{\rm HP}$, while that of the the
extremal superentropic Kerr-Newman-AdS$_4$ black hole always exceeds the same scale
$r_{\rm HP}$. This might be taken as a direct signature to distinguish these two extremal
ultraspinning charged AdS$_4$ black holes, which hints that it is possible to judge
them via the observation of their shadow sizes.
\end{enumerate}

There are two promising further topics to be pursued in the future. As mentioned above,
one intriguing topic is to explore the black hole shadows of two ultraspinning charged
AdS$_4$ solutions, and this might shed light on our knowledge of black holes in gauged
supergravity theories. And the other is to extend the present work to the more general
dyonic case. We hope to report the related progress along these directions soon.

\acknowledgments

This work is supported by the National Natural Science Foundation of China under Grants
No. 11775077, No. 11690034, No. 11435006, No. 11675130, and No. 11275157, and by the
Science and Technology Innovation Plan of Hunan province under Grant No. 2017XK2019.

\textit{Notes added}.---After submission of this paper for publication, we have reconsidered
the work, in particular about the bound on the horizon radius of extremal ultraspinning
Kerr-Sen-AdS$_4$ black hole by making a shift on the radial coordinate $r = \rho -b$,
so that the metric looks like more similar to the superentropic Kerr-Newman-AdS$_4$
solution as now we have the structure function $\breve{\Delta}_\rho = (\rho^2 +l^2
-b^2)^2l^{-2} -2m\rho +q^2$. Repeating the computations as did in the Sec. \ref{bounds},
we obtain again $m_e \geqslant 8l/\sqrt{27}$ but now $\rho_e \geqslant l/\sqrt{3}$,
and all other results of this paper remain unchanged. \footnote{Of course, one should
correspondingly replace $r_+$ by $\rho_+ -b$ in all the expressions that appeared in the
main context.} This can be easily understood as now one has $\rho_e = 2b_e/3
+\sqrt{b_e^2 +3l^2}/3 \geqslant l/\sqrt{3}$ due to $0 \leqslant b_e < \infty$.
That is to say, when a different radial coordinate is adopted, both the extremal
ultraspinning Kerr-Sen-AdS$_4$ black hole and the superentropic Kerr-Newman-AdS$_4$
black hole have the same lower bound on the horizon radius. Accordingly this means that
one might not discriminate them via observing their shadow sizes.

\end{document}